\documentclass[twocolumn,showpacs,preprintnumbers,amsmath,amssymb]{revtex4}
\usepackage{graphicx}
\usepackage{dcolumn}
\usepackage{color}
\usepackage{bm}
\begin{document}
\title {Landau-Drude Diamagnetism: Fluctuation, Dissipation and Decoherence.}
\vskip -0.5cm \author{ Malay Bandyopadhyay$^{1}$ and Sushanta Dattagupta$^{1,2}$}
\affiliation{$^1$S.N. Bose National Centre for Basic Sciences, JD Block, Sector III, Salt Lake, Kolkata 700098, India.\\
$^2$Jawaharlal Neheru Centre for Advanced Scientific Research, Jakkur, Bangalore 560064, India.}

\date{\today}
\begin{abstract}
Starting from a quantum Langevin equation (QLE) of a charged particle coupled to a heat bath in the presence of an external magnetic field, we present a fully dynamical calculation of the susceptibility tensor. We further evaluate the position autocorrelation function by using the Gibbs ensemble approach. This quantity is shown to be related to the imaginary part of the dynamical susceptibility, thereby validating the fluctuation-dissipation theorem in the context of dissipative diamagnetism. Finally we present an overview of coherence-to-decoherence transition in the realm of dissipative diamagnetism at zero temperature. The analysis underscores the importance of the details of the relevant physical quantity, as far as coherence to decoherence transition is concerned.  
\end{abstract}
\maketitle
{\section {Introduction}}
The problem of a quantum charged particle in the presence of a magnetic field is an old and important one \cite{sakurai}. When Landau gave the theory of diamagnetism a major breakthrough in solid state physics was made possible \cite{landau,van vleck}. The physics of Landau levels is of great interest in the quantum Hall effect \cite{imry} and high temperature superconductivity \cite{hong}. In the present paper we address the issue of what happens when we combine the Landau problem with the Drude transport treatment which naturally brings in the phenomenon of environment-induced dynamics \cite{ashcroft}.\\
The consequences of coupling of a system to its environment are threefold. First, energy may irreversibly be transferred from the system to the environment giving rise to dissipation \cite{legg1,legg2,legg3}. Second, the spontaneous fluctuations injected by the environment into the system govern the response of the system degrees of freedom to weak, external stimuli \cite{weiss,hanggi}. Finally the entanglement between the system and environment degrees of freedom destroys the coherent superposition of quantum states, leading to decoherence \cite{zurek}.\\
We discuss all the three above mentioned effects in the context of Landau diamagnetism which is inherently and intrinsically quantum in nature. For the purpose of investigating fluctuation, dissipation and decoherence in what we call Landau-Drude diamagnetism \cite{sdg} it is convenient to use the formulation given by Ford et al \cite{ford1,ford2}, following the classical treatment due to Zwanzig \cite{zwan}. Starting from the Feynmann-Vernon model in which a particle moving in an arbitrary potential is assumed to be linearly coupled with a collection of quantum harmonic oscillators \cite{feyn}, these authors derive a quantum Langevin equation (QLE). We use this QLE as the basis of our further discussion, in what may be referred to as the Einstein approach to Statistical Physics \cite{kadanoff}.\\ 
From QLE we calculate the dynamical susceptibility tensor. We then evaluate the position autocorrelation function from the Euclidean action of the Feynmann-Vernon model using the Gibbsian ensemble approach. This quantity is a measure of spontaneous fluctuations in the degrees of freedom of the system. We discuss the relation between the position autocorrelation function and the imaginary part of the susceptibility and thus establish the fluctuation-dissipation theorem in the context of dissipative diamagnetism.\\
The destruction of quantum coherence by environment-induced dissipation is of central interest in atomic physics \cite{garraway}, condensed matter physics \cite{grabert}, as well as chemical and biological reactions \cite{chandler}. We discuss this environment-induced decoherence in the context of dissipative diamagnetism. Landau diamagnetism has its origin in coherent circular motion of the electron in a plane normal to the magnetic field. This coherent motion is disturbed due to interaction with environmental degrees of freedom, e.g. defects, phonons, etc. We illustrate how the system transits from the coherent `Landau regime' to the decoherent `Bohr-Van Leeuwen regime' \cite{bohr,van}. We show that the coherent-decoherent transition depends on the particular dynamical quantity (e.g., correlation function, occupation probability, etc.) under consideration.\\
The paper is organized as follows. In the next section we discuss our model Hamiltonian and the corresponding QLE. In Sec. III we calculate the generalized susceptibility tensor. Sec. IV deals with the position autocorrelation function and its relation to the susceptibility, thus establishing the fluctuation-dissipation theorem in the context of dissipative diamagnetism. In Sec. V we study the coherence-to-decoherence transition. Finally we summarize our results and present a few concluding remarks in Sec.VI.\\
{\section {Model, QLE and Einstein Approach}}
We start with the Feynmann-Vernon Hamiltonian for a charged particle in a magnetic field ${\vec{B}}$, coupled to an environment of quantum harmonic oscillators  \cite{feyn}
\begin{eqnarray}
{\cal H} & = & \frac{1}{2m}\Big({\vec{p}}-\frac{e}{c}{\vec{A}}\Big)^2 +\frac{1}{2}m\omega_0^2{\vec {q}}^2
\nonumber \\
& & + \sum_j\Big\lbrack\frac{1}{2m_j}{\vec {p_j}}^2 + \frac{1}{2}m_j\omega_j^2({\vec{q}}_j - {\vec{q}}\Big)^2\Big\rbrack,
\end{eqnarray}
where $\vec{p}$ and $\vec{q}$ are the momentum and position operators of the particle, and $\vec{A}$ is the vector potential. The second term due to Darwin \cite{darwin} represents a confining potential to recover the correct boundary contribution. Now following Ford et al \cite{ford1,ford2} one can write the QLE emanating from Eq.(1) as \cite{sdg}
\begin{equation}
m\ddot{\vec {q}} + \int_{-\infty}^{t}dt^{\prime}\gamma(t-t^{\prime})\dot{\vec{q}}(t^{\prime}) + m\omega_0^2\vec{q} - \frac{e}{c}(\dot{\vec{q}}\times\vec{B}) = \vec{F}(t),
\end{equation}
where the auto-correlation and the commutator of $\vec{F}(t)$ are given by
\begin{eqnarray}
\Big{<}\{F_{\alpha}(t),F_{\beta}(t^{\prime})\}\Big{>} = \delta_{\alpha\beta}\frac{2}{\pi}\int_0^{\infty}\Re\lbrack{\tilde{\gamma}}(\omega+i0^+)\rbrack \hbar\omega
\nonumber \\
\coth(\frac{\hbar\omega}{2k_BT})\cos\lbrace\omega(t-t^{\prime})\rbrace d\omega, 
\end{eqnarray}
\begin{eqnarray}
\Big{<}\lbrack F_{\alpha}(t),F_{\beta}(t^{\prime})\rbrack\Big{>} = \delta_{\alpha\beta}\frac{2}{i\pi}\int_0^{\infty}\Re\lbrack{\tilde{\gamma}}(\omega+i0^+)\rbrack\hbar\omega
\nonumber \\
\sin\lbrace\omega(t-t^{\prime})\rbrace d\omega, 
\end{eqnarray}
where $\tilde{\gamma}(s) = \int_{0}^{\infty}dt\exp(ist)\gamma(t)$ ($Im$ s $> 0$).\\
 At this stage we introduce the nomenclature of Ohmic dissipation as well as non-Ohmic dissipation. Defining the spectral density of the environmental degrees of freedom as $J(\omega) = \frac{\pi}{2}\sum_{j=1}^{N}m_j\omega_j^3\delta(\omega-\omega_j)$, we can rewrite the memory kernel $\gamma(t)$ in terms of the spectral density as
\begin{equation}
\gamma(t) = \Theta(t)\frac{2}{m\pi}\int_0^{\infty}d\omega\frac{J(\omega)}{\omega}\cos(\omega t),
\end{equation}
where $\Theta(t)$ is the Heaviside step function. In the strictly Ohmic case, damping is frequency-independent and the spectral density $J(\omega) = m\gamma\omega$. The memory kernel $\gamma(t-t^{\prime})$ is thus replaced by $m\gamma\delta(t-t^{\prime})$, so that $\Re\lbrack\tilde{\gamma}(\omega+i0^+)\rbrack$ reduces to $m\gamma$, a constant. In this limit we get an ordinary Langevin equation. It is interesting to note that the underlying stochastic process is still non-Markovian, even though there is no memory. In the non-Ohmic case (e.g. due to a phonon bath) the spectral density is defined as $J(\omega) = m\tilde{\gamma}(\omega)$, where $\tilde{\gamma}(\omega)=\gamma\omega^3$. The damping kernel $\tilde{\gamma}(\omega)$ then brings in memory-friction effects.\\   
{\section {Generalized Susceptibility Tensor}}
In this section we consider the linear response of the position coordinate to an external force $\vec{f}(t)$, assumed small. By imagining the force to have been switched on at time $t = -\infty$ all transient effects can be ignored and the nontransient response can be captured by the frequency-dependent generalised susceptibility. The corresponding QLE now reads
\begin{eqnarray}
m\ddot{\vec {q}}+\int_{-\infty}^{t}dt^{\prime}\gamma(t-t^{\prime})\dot{\vec{q}}(t^{\prime})+m\omega_0^2\vec{q} - \frac{e}{c}(\dot{\vec{q}}\times\vec{B}) = 
\nonumber \\
\vec{F}(t)+\vec{f}(t).
\end{eqnarray}
We rewrite Eq. (6) in a Fourier transformed form
\begin{eqnarray}
\Big\lbrack (m(\omega_0^{2}-\omega^2)-i\omega\tilde{\gamma}(\omega))\delta_{\alpha\beta}+i\omega\frac{e}{c}\epsilon_{\alpha\beta\rho} B_{\rho}\Big\rbrack\tilde{q}_{\beta}(\omega) =
\nonumber \\
\tilde{F}_{\alpha}(\omega)+\tilde{f}_{\alpha}(\omega),
\end{eqnarray}
with
\begin{eqnarray*}
\tilde{Z}_i(\omega) & = & \int_0^{\infty}dt e^{i\omega t}Z_i(t)\\
& & (i=1,2,3,4; Z_1=\gamma, Z_2=q_{\beta}, Z_3 = F_{\alpha}, Z_4 = f_{\alpha}),
\end{eqnarray*}
and $\epsilon_{\alpha\beta\rho}$ is the Levi-Civita symbol, $\alpha,\beta, \rho$ being the three spatial directions (i.e. $\alpha,\beta,\rho$ = x, y, z). In linear response theory one can write \cite{sdg3} 
\begin{equation}
q_{\alpha}(t) = \int_{-\infty}^{t}ds \chi_{\alpha\beta}(t-s)(F_{\beta}(s)+f_{\beta}(s)), 
\end{equation}
where $\chi_{\alpha\beta}$ is the generalised susceptibility tensor. In Fourier transformed form Eq. (8) becomes
\begin{equation}
\tilde{q}_{\alpha}(\omega) = \chi_{\alpha\beta}(\omega)\lbrack\tilde{F}_{\beta}(\omega)+\tilde{f}_{\beta}(\omega)\rbrack.
\end{equation}
\begin{figure}[t!]
{\rotatebox{0}{\resizebox{7cm}{5.75cm}{\includegraphics{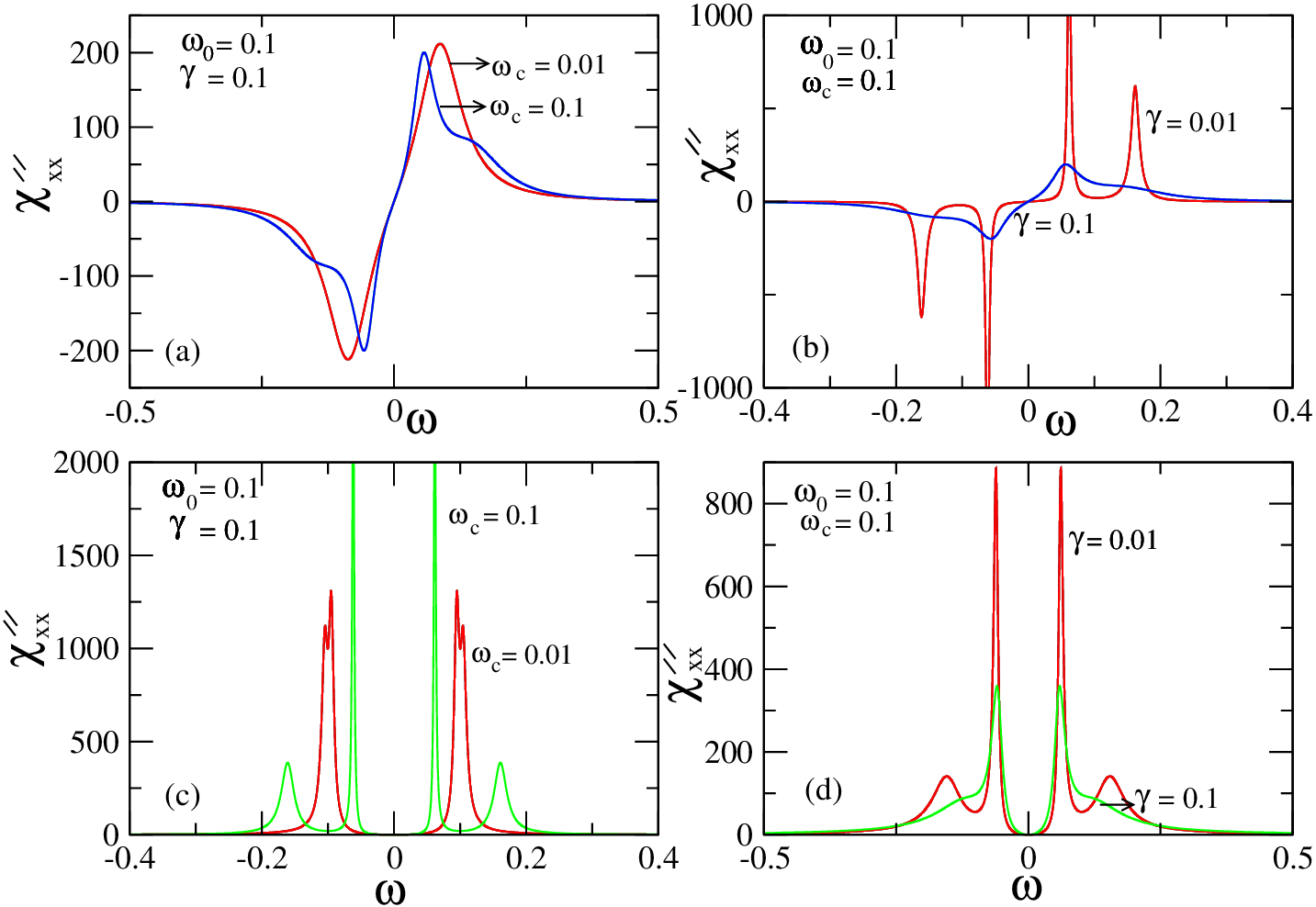}}}}
\caption{(color online) The imaginary part of susceptibility $\chi_{xx}$ (a) Ohmic dissipation case $(J(\omega)\sim\omega)$ for two $\omega_c$ values. (b) Ohmic dissipation case for two $\gamma$ values. (c) Non-Ohmic dissipation case  $(J(\omega)\sim\omega^3)$ for two $\omega_c$ values. (d) Non-Ohmic dissipation case  for two $\gamma$ values.}
\end{figure}
Rewriting Eq. (7) as 
\begin{equation}
Y_{\alpha\beta}(\omega)\tilde{q}_{\beta}(\omega) = \lbrack\tilde{F}_{\alpha}(\omega)+\tilde{f}_{\alpha}(\omega)\rbrack,
\end{equation}
the generalised susceptibility can be evaluated as
\begin{equation}
\chi_{\alpha\beta} = \lbrack Y^{-1}(\omega)\rbrack_{\alpha\beta},
\end{equation} 
with \begin{eqnarray} 
Y(\omega) = \left( \begin{array}{ccc}
\Delta(\omega) & i\omega\frac{e}{c}B_{z} & -i\omega\frac{e}{c}B_{y}\\ 
-i\omega\frac{e}{c}B_{z} &\Delta(\omega)  &i\omega\frac{e}{c}B_{x}\\ 
i\omega\frac{e}{c}B_{y} & -i\omega\frac{e}{c}B_{x} &\Delta(\omega)   \end{array} \right),
\end{eqnarray}
where $\Delta(\omega) = m(\omega_0^2 - \omega^2) - i\omega\tilde{\gamma}(\omega)$. Clearly
\begin{eqnarray}
\chi(\omega) =\frac{1}{Det\lbrack Y(\omega)\rbrack}\left( \begin{array}{ccc}\chi_{xx} &\chi_{xy}&\chi_{xz} \\
\chi_{yx}&\chi_{yy}&\chi_{yz}\\
\chi_{zx}&\chi_{zy}&\chi_{zz}\end{array} \right),
\end{eqnarray}
where 
\begin{eqnarray}
\hskip -0.5cm Det\lbrack Y(\omega)\rbrack & = &\Delta(\omega)\lbrack\Delta^2(\omega)-(\omega\frac{e}{c})^2\vec{B}^2\rbrack;
\nonumber \\
\chi_{ii} & = &\Delta^2(\omega) - (\omega\frac{e}{c})^2B_i^2, (i = x,y,z);
\nonumber \\
\chi_{xy} & = &\chi_{yx}^{*} = -(\omega\frac{e}{c})^2B_{x}B_{y}-i\omega\frac{e}{c}B_{z}\Delta(\omega);
\nonumber \\
\chi_{xz} & = &\chi_{zx}^{*}  =  -(\omega\frac{e}{c})^2B_{x}B_{z}+i\omega\frac{e}{c}B_{y}\Delta(\omega);
\nonumber \\  
\chi_{yz} & = & \chi_{zy}^{*}  = -(\omega\frac{e}{c})^2B_{y}B_{z}-i\omega\frac{e}{c}B_{\alpha}\Delta(\omega),
\end{eqnarray}
where $(*)$ denotes the complex conjugate of the corresponding variable.
\begin{figure}[t!]
{\rotatebox{0}{\resizebox{7cm}{8cm}{\includegraphics{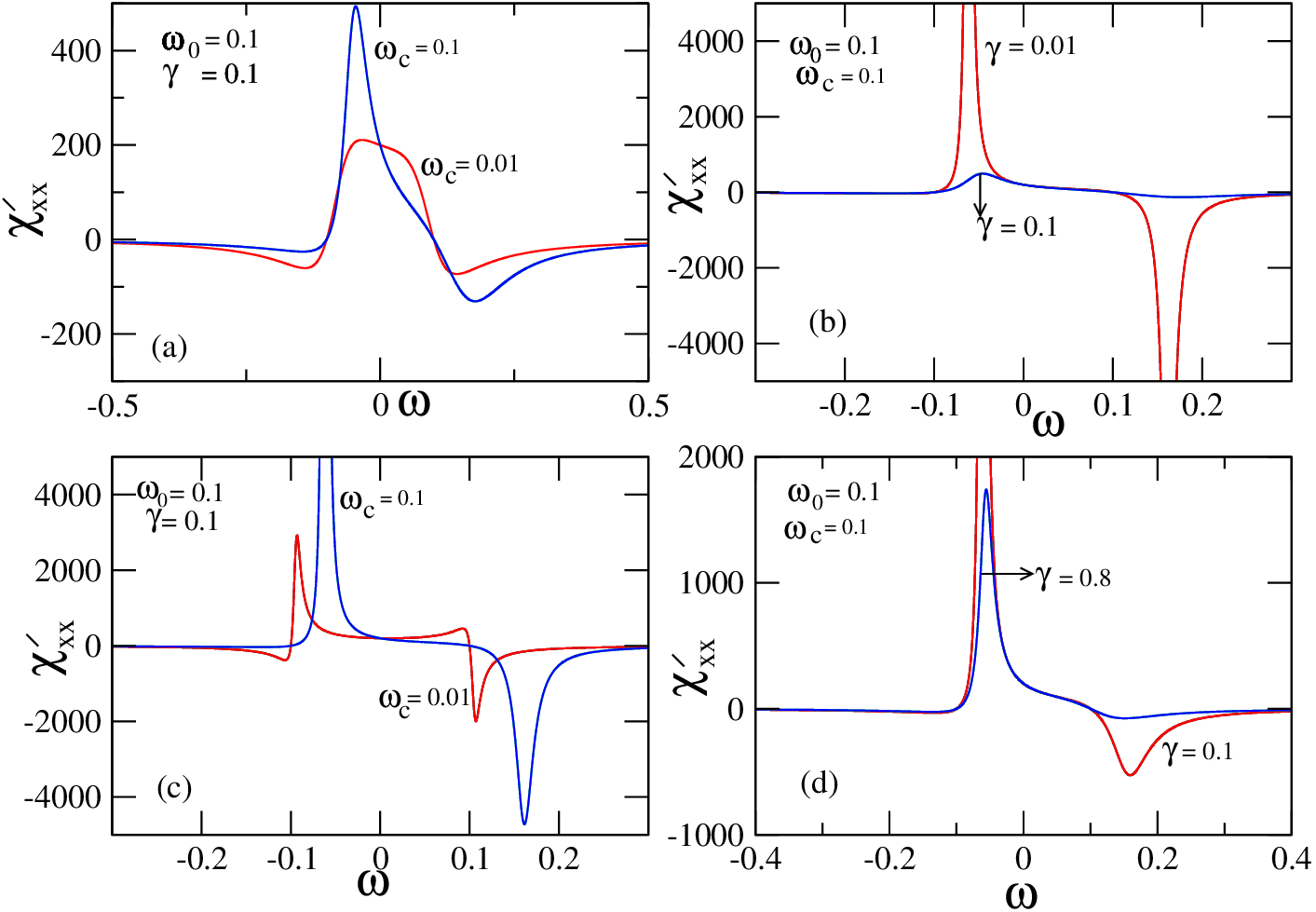}}}}
\caption{(color online) The real part of susceptibility $\chi_{xx}$ (a) Ohmic dissipation case $(J(\omega)\sim\omega)$ for two $\omega_c$ values. (b) Ohmic dissipation case for two $\gamma$ values. (c) Non-Ohmic dissipation case  $(J(\omega)\sim\omega^3)$ for two $\omega_c$ values. (d) Non-Ohmic dissipation case for two $\gamma$ values.}
\end{figure}

The expression is simplified when the magnetic field is taken along z axis, thus
\begin{widetext}
\begin{eqnarray}
\chi(\omega) =\frac{1}{Det\lbrack Y(\omega)\rbrack}\left( \begin{array}{ccc}
\Delta^2(\omega)&-i\omega\frac{e}{c}\Delta(\omega)B & 0 \\
i\omega\frac{e}{c}\Delta(\omega)B &\Delta^2(\omega) & 0\\
 0 & 0 &\Delta^2(\omega)-(\omega\frac{e}{c})^2B^2 \end{array} \right).
\end{eqnarray}
\end{widetext}
For this particular case the real part of the susceptibility is
\begin{eqnarray}
 \chi_{xx}^{\prime}& = &\chi_{yy}^{\prime} 
\nonumber \\
&=& \frac{1}{2m^2}\Big\lbrack\frac{(\omega_0^2-\omega^2+\omega\omega_c/2)}{(\omega^2-\omega_0^2+\omega\omega_c)^2+\frac{\omega^2\tilde{\gamma}^{2}(\omega)}{m^2}}
\nonumber \\ 
& & +\frac{(\omega_0^2-\omega^2-\omega\omega_c/2)}{(\omega^2-\omega_0^2-\omega\omega_c)^2+\frac{\omega^2\tilde{\gamma}^{2}(\omega)}{m^2}}\Big\rbrack, 
\end{eqnarray}
and the imaginary part is
\begin{eqnarray}
\chi_{xx}^{\prime\prime} & = & \chi_{yy}^{\prime\prime} 
\nonumber \\
&=& \frac{\tilde{\gamma}(\omega)\omega}{2m^2}\Big\lbrack\frac{1}{(\omega^2-\omega_0^2+\omega\omega_c)^2+\frac{\omega^2\tilde{\gamma}^{2}(\omega)}{m^2}}
\nonumber \\
& & +\frac{1}{(\omega^2-\omega_0^2-\omega\omega_c)^2+\frac{\omega^2\tilde{\gamma}^{2}(\omega)}{m^2}}\Big\rbrack, 
\end{eqnarray}
where the cyclotron frequency $\omega_c = \frac{eB}{mc}$.
For the Ohmic dissipation case the susceptibility  has four poles at 
\begin{eqnarray}
\omega = \pm\tilde{\omega}_+ = \lbrack\frac{\omega_c+i\gamma}{2}\pm \frac{\sqrt{4\omega_0^2+\omega_c^2-\gamma^2+2i\omega_c\gamma}}{2}\rbrack
\nonumber \\ 
\omega = \pm \tilde{\omega}_- =  \lbrack\frac{-\omega_c+i\gamma}{2} \pm \frac{\sqrt{4\omega_0^2+\omega_c^2-\gamma^2-2i\omega_c\gamma}}{2}\rbrack.
\end{eqnarray} 
On the other hand, for the non-Ohmic case these poles cannot be evaluated analytically. The numerical results for the Ohmic dissipation as well as non-Ohmic dissipation cases are presented below.\\
We plot in Fig. (1) the dissipative part of the x-component of susceptibility i.e. $\chi^{\prime\prime}_{xx}(\omega)$ versus $\omega$ for different values of $\omega_c$ and $\gamma$ in accordance with Eq. (17). We note that $\chi^{\prime\prime}_{xx}(\omega)$ is odd in $\omega$ for the Ohmic dissipation case and has Lorentzian line shapes for finite damping values, with peaks centered at the poles. For the non-Ohmic case $\chi^{\prime\prime}_{xx}(\omega)$ is even in $\omega$. It is evident from Fig. (1-b) that for finite but weak damping one can obtain all the four peaks for the ohmic dissipation case whereas for high damping only two peaks are obtained. The same is true for the non-Ohmic case (Fig. (1-d)). The only difference is that the magnitude of the peak height is higher for the non-ohmic case and is always positive. Also the peak width increases with the increase of $\gamma$ for both Ohmic and non-Ohmic cases. On the other hand the width of the peak decreases with the increase of $\omega_c$, as is expected on physical grounds. In the non-Ohmic case the number of peaks also increases from two to four with the increase of $\omega_c$, whereas it remains two for the Ohmic case with the increase of $\omega_c$, if $\gamma$ is kept large. Thus, dissipative effects are stronger for the Ohmic case.\\
In Fig. (2) we plot the reactive part or the real part of the x-component of susceptibility ($\chi^{\prime}_{xx}(\omega)$) versus $\omega$ for different values of $\omega_c$ and $\gamma$ in accordance with Eq. (16). $\chi^{\prime}_{xx}(\omega)$ is odd in $\omega$ for the Ohmic as well as non-Ohmic cases.  The spreading of the peaks increases but the peak height decreases with the decrease of $\omega_c$ for the Ohmic case. On the other hand both the spreading and peak height decrease with the decrease of $\omega_c$ for the non-Ohmic case. But the features are same with the variation of $\gamma$ for both Ohmic and non-Ohmic cases ---- the peak height increases but the spreading decreases with the decrease of $\gamma$. In addition the number of peaks increases from one to two with the decrease of $\gamma$ in the Ohmic as well as non-Ohmic cases.\\ 
The z component of the susceptibility tensor is of course the same as that of a damped harmonic oscillator because it has nothing to do with $\omega_c$.\\
{\section {Fluctuation - Dissipation Relationship: Gibbs Approach}}
In Sec. III we calculated the susceptibility as the asymptotic (i.e. $t\rightarrow\infty$) response from a fully time-dependent formulation of the underlying QLE. Because detailed balance relations (viz. Eqs. (3) and (4)) are built-in within the QLE, as the heat bath is assumed to be in thermal equilibrium at a fixed temperature T, the asymptotic response is expected to be related to the equilibrium properties of the system. This expectation is at the heart of what Kadanoff calls the Einstein approach to Statistical Mechanics \cite{kadanoff} in which equilibrium answers are sought to be obtained from the asymptotic limit of time-dependent results. It is then natural to ask whether the response obtained from the Einstein approach can be related to spontaneous or equilibrium fluctuations that can be independently calculated from the standard Gibbsian formulation of equilibrium Statistical Mechanics. If we can establish this relation it will not only be tantamount to establishing the fluctuation-dissipation theorem for the phenomena at hand, but also to demonstrating the equivalence of the Einstein and the Gibbs approaches to Statistical Mechanics \cite{malay}.\\
 With this preamble the position autocorrelation function in equilibrium is defined as
\begin{equation}
C(t) = <\vec{x}(t)\cdot\vec{x}(0)> = Tr(\vec{x}(t)\cdot\vec{x}(0)\rho_{\beta}),
\end{equation}
where $\rho_{\beta}$ is the equilibrium density matrix of the full system and $\vec{x}$ is the two dimensional position vector in the x-y plane.. We determine $C(t)$ by first calculating its imaginary time version starting from the  Euclidean action of the system as described by Eq. (1)
\begin{eqnarray}
S^E[\vec{x}] & = & \int_0^{\hbar\beta}d\tau\Big(\frac{m}{2}\dot{\vec{x}}^2+\frac{m}{2}\omega_0^2\vec{x}^2+im\omega_c(\dot{\vec{x}}\times\vec{x})_z\Big)  
\nonumber \\
& & +\frac{1}{2m}\int_0^{\hbar\beta}d\tau \int_0^{\hbar\beta}d\sigma \tilde{\gamma}(\tau-\sigma)\vec{x}(\tau)\cdot\vec{x}(\sigma)
\nonumber \\
& & +\int_0^{\hbar\beta}d\tau \vec{f}(\tau)\cdot\vec{x}(\tau),
\end{eqnarray}
where the first term takes care of the system part, the second term accounts for the coupling to the environment and the third term corresponds to the interaction with an external force in imaginary time. This helps us to determine the correlation function by variation with respect to this force \cite{ingold1,ingold2}
\begin{equation}
<\vec{x}(\tau)\cdot\vec{x}(\sigma)> = \hbar^2Tr\Big(\frac{\delta}{\delta \vec{f}(\tau)}\frac{\delta}{\delta \vec{f}(\sigma)}\rho_{\beta}\Big)_{\vec{f}=0}.
\end{equation}
\begin{figure}[!b]
{\rotatebox{0}{\resizebox{5cm}{3.5cm}{\includegraphics{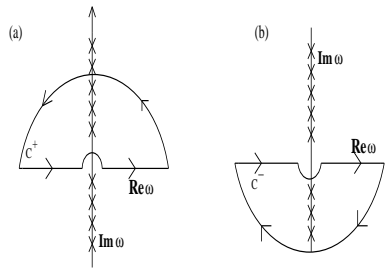}}}}
\caption{The analytic continuation of the imaginary time correlation function to real times by using the contours depicted in (a) and (b) to obtain Eq. (24) and Eq. (25) respectively.}
\end{figure}
It is sufficient to restrict ourselves to the classical path for the calculation of the autocorrelation function \cite{ingold1,ingold2}. Thus the Fourier representation of the classical Euclidean action becomes \cite{malay,ingold1} 
\begin{eqnarray}
\hskip -0.8cm S_{cl}^E & = & -\frac{1}{2m\hbar\beta}\sum_{n=-\infty}^{+\infty}\Big\lbrack\frac{1}{\nu_n^2+\frac{\tilde{\gamma}(|\nu_n|)\nu_n}{m}+\omega_0^2-i\omega_c\nu_n}
\nonumber \\
& & +\frac{1}{\nu_n^2+\frac{\tilde{\gamma}(|\nu_n|)\nu_n}{m}+\omega_0^2+i\omega_c\nu_n}\Big\rbrack
\nonumber \\
& &\times \int_0^{\hbar\beta}d\tau\int_0^{\hbar\beta}d\sigma \vec{f}(\tau)\vec{f}(\sigma)\exp(i\nu_n(\tau-\sigma)),
\end{eqnarray}
where $\nu_n = \frac{2\pi n}{\hbar\beta}$ are the so-called Matsubara frequencies. Since the force appears only through the action in the exponent of the equilibrium density matrix, we can easily evaluate the functional derivatives according to Eq. (21) and obtain the position autocorrelation function in imaginary time:
\begin{eqnarray}
\hskip -0.5cm C(\tau)& = & \frac{1}{m\beta}\sum_{n=-\infty}^{+\infty}\Big\lbrack\frac{1}{\nu_n^2+\frac{\tilde{\gamma}(|\nu_n|)\nu_n}{m}+\omega_0^2-i\omega_c\nu_n}
\nonumber \\
& &+\frac{1}{\nu_n^2+\frac{\tilde{\gamma}(|\nu_n|)\nu_n}{m}+\omega_0^2+i\omega_c\nu_n}\Big\rbrack\exp(i\nu_n\tau).
\end{eqnarray}
The real time correlation function cannot be obtained by simply replacing $\tau$ by $ it$, because for negative times the sum does not converge. The idea is to express the sum in Eq. (23) as a contour integral in the complex frequency plane. To do this, we need a standard trick of Statistical Mechanics \cite{reif}. Look for a function which is well-behaved at infinity, but has poles at $\omega = i\nu_n$. This requirement is fulfilled by $\frac{\hbar\beta}{1-\exp(-\hbar\beta\omega)}$. Now doing the integration along the contour shown in Fig. (3-a) we find
\begin{widetext}
\begin{eqnarray}
& &\int_{C^+}d\omega\frac{\hbar\beta}{1-\exp(-\hbar\beta\omega)}\Big\lbrack\frac{\exp(-\omega\tau)}{\omega_0^2-\omega^2+\frac{i\tilde{\gamma}(\omega)\omega}{m} + \omega\omega_c}   
+\frac{\exp(-\omega\tau)}{\omega_0^2-\omega^2+\frac{i\tilde{\gamma}(\omega)\omega}{m} - \omega\omega_c}\Big\rbrack 
\nonumber \\
& = & -2i\frac{\pi}{\omega_0^2}-2\pi i\Big\lbrack \sum_{n=1}^{\infty}\frac{\exp(i\nu_n\tau)}{\nu_n^2+\frac{\tilde{\gamma}(|\nu_n|)\nu_n}{m}+\omega_0^2-i\omega_c\nu_n} 
+\sum_{n=1}^{\infty}\frac{\exp(i\nu_n\tau)}{\nu_n^2+\frac{\tilde{\gamma}(|\nu_n|)\nu_n}{m}+\omega_0^2+i\omega_c\nu_n}\Big\rbrack.
\end{eqnarray}
Similarly, an integration along the contour shown in Fig. (3-b) leads to
\begin{eqnarray}
& & \int_{C^-}d\omega\frac{\hbar\beta}{1-\exp(-\hbar\beta\omega)}\Big\lbrack\frac{\exp(-\omega\tau)}{\omega_0^2-\omega^2-\frac{i\tilde{\gamma}(\omega)\omega}{m} - \omega\omega_c}  
 +\frac{\exp(-\omega\tau)}{\omega_0^2-\omega^2-\frac{i\tilde{\gamma}(\omega)\omega}{m} + \omega\omega_c}\Big\rbrack
\nonumber \\
& = & 2i\frac{\pi}{\omega_0^2}+2\pi i \Big\lbrack\sum_{n=-\infty}^{-1}\frac{\exp(i\nu_n\tau)}{\nu_n^2+\frac{\tilde{\gamma}(|\nu_n|)\nu_n}{m}+\omega_0^2-i\omega_c\nu_n} 
+\sum_{n=-\infty}^{-1}\frac{\exp(i\nu_n\tau)}{\nu_n^2+\frac{\tilde{\gamma}(|\nu_n|)\nu_n}{m}+\omega_0^2+i\omega_c\nu_n}\Big\rbrack.
\end{eqnarray}
Subtracting Eq. (25) from Eq. (24) we obtain
\begin{eqnarray}
& &\frac{1}{m\beta}\sum_{n=-\infty}^{+\infty}\Big\lbrack\frac{1}{\nu_n^2+\frac{\tilde{\gamma}(|\nu_n|)\nu_n}{m}+\omega_0^2-i\omega_c\nu_n} 
 +\frac{1}{\nu_n^2+\frac{\tilde{\gamma}(|\nu_n|)\nu_n}{m}+\omega_0^2+i\omega_c\nu_n}\Big\rbrack e^{(i\nu_n\tau)} 
\nonumber \\
& = &\frac{\hbar}{m^2\pi}\int_{-\infty}^{+\infty}d\omega \tilde{\gamma}(\omega)\omega\Big\lbrack\frac{1}{(\omega^2-\omega_0^2-\omega\omega_c)^2+\frac{\tilde{\gamma}^2(\omega)\omega^2}{m^2}} 
+\frac{1}{(\omega^2-\omega_0^2+\omega\omega_c)^2+\frac{\tilde{\gamma}^2(\omega)\omega^2}{m^2}}\Big\rbrack\frac{e^{(-\omega\tau)}}{1-e^{(-\hbar\beta\omega)}}.
\end{eqnarray} 
We may now pass to real time by the replacement $\tau\rightarrow it$ to obtain the real time correlation function
\begin{eqnarray}
C(t) = \frac{\hbar}{\pi m^2}\int_{-\infty}^{+\infty}d\omega \Big\lbrack\frac{\tilde{\gamma}(\omega)\omega}{(\omega^2-\omega_0^2-\omega\omega_c)^2+\frac{\tilde{\gamma}^2(\omega)\omega^2}{m^2}}
+\frac{\tilde{\gamma}(\omega)\omega}{(\omega^2-\omega_0^2+\omega\omega_c)^2+\frac{\tilde{\gamma}^2(\omega)\omega^2}{m^2}}\Big\rbrack\frac{e^{(-i\omega t)}}{1-e^{(-\hbar\beta\omega)}}. 
\end{eqnarray}
\end{widetext}
It is easy to show from Eq. (27) that
\begin{equation}
\tilde{C}(\omega) = \frac{2\hbar}{1-\exp(-\beta\hbar\omega)}\chi_{xx}^{\prime\prime}(\omega).
\end{equation}
Eq. (28) represents the fluctuation-dissipation theorem in the context of dissipative Landau diamagnetism. The position autocorrelation function describes the spontaneous fluctuations of the system while the imaginary part of the dynamic susceptibility $\chi_{xx}^{\prime\prime}$ determines the energy dissipation in the system due to work done by an external weak force.\\
{\section {Coherence - Decoherence transition}}
In this Section our discussion is focused on the destruction of quantum coherence by environment-induced dissipation in the context of dissipative diamagnetism. Two questions are relevant: (i) Can we quantify the criterion for crossover from coherent to decoherent dynamics? (ii) Is this criterion universal? As far as some model systems are concerned, the answer to (i) is in the affirmative \cite{egger}. Regarding the question (ii), there seems to be no universality in the criterion of crossover. As a matter of fact, the value of the crossover parameter depends on the particular quantity under consideration and its initial preparation. Thus, quantum memory effects play a crucial role as the system makes a transition from the coherent to the decoherent regime. To clarify this issue we focus on dissipative diamagnetism and we consider its $T=0$ behavior, wherein quantum coherence is the most prominent. Here we follow the discussion of Egger et al \cite{egger}.\\
We start with the QLE for dissipative Landau diamagnetism subject to Ohmic damping. The motion in the x-y plane can be expressed in the compact form:
\begin{equation}
\ddot{Z}+\bar{\gamma}\dot{Z}+\omega_0^2Z = \frac{\theta(t)}{m},
\end{equation}
where $Z = x+iy$, $\bar{\gamma}=\gamma+i\omega_c$, and $\theta = F_x +iF_y$. Thus, the time-dependence of the corresponding classical quantity ({\em a la} Ehrenfest) is governed by the following equation:
\begin{equation}
<\ddot{Z}> +\bar{\gamma}<\dot{Z}> +\omega_0^2<Z> = \frac{\theta(t)}{m},
\end{equation}
where the angular brackets represent statistical averages over the ground state properties ($T=0$), i.e. the expectation values. As discussed earlier the response to an external force is characterized by the generalized susceptibility $\chi_{osc}(t)$ \cite{sdg3}:
\begin{equation}
<Z(t)> = \frac{1}{m\omega_0}\int_{-\infty}^{t}dt^{\prime}\chi_{osc}(t-t^{\prime})\theta(t^{\prime}).
\end{equation}

From Eqs. (24) and (25), we obtain the Fourier transform of $\chi_{osc}(t)$ as
\begin{equation}
\chi_{osc}(\omega) = \frac{\omega_0}{\omega_0^2-\omega^2-i\bar{\gamma}\omega}.
\end{equation}
On the other hand, using the fluctuation-dissipation theorem \cite{sdg3}, $\chi_{osc}(\omega)$ can be related to the spectral function $S_{osc}(\omega)$, which in turn determines the equilibrium correlation function $C_{osc}(\omega)$. The functional relationship which holds at $T = 0$ is as follows:
\begin{equation}
\Im \chi_{osc}(\omega)=\omega S_{osc}(\omega) = \frac{\omega}{|\omega|}C_{osc}(\omega),
\end{equation} 
where $C_{osc}(t) = \Re<Z(t)Z(0)>$. Using Eqs. (32) and (33), we obtain the spectral function
\begin{equation}
S_{osc}(\omega) = \frac{\gamma \omega_0}{(\omega_0^2-\omega^2+\omega\omega_c)^2+\gamma^2\omega^2}.
\end{equation} 
\begin{figure}[t!]
{\rotatebox{0}{\resizebox{8cm}{8cm}{\includegraphics{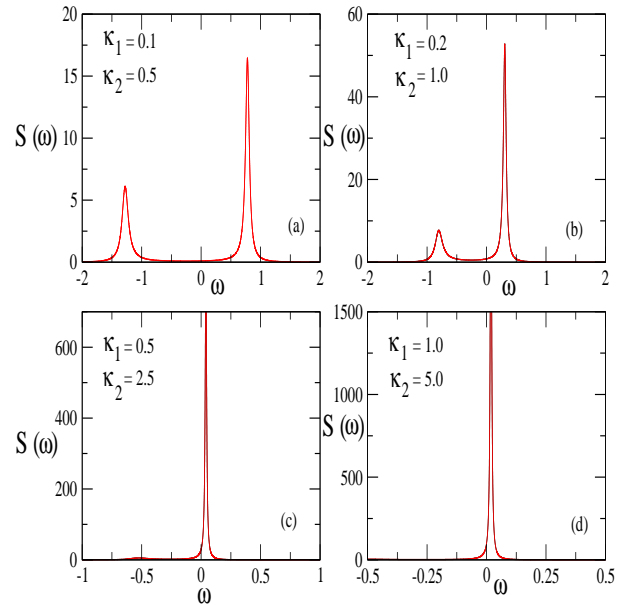}}}}
\caption{(color online) Spectral function $S_{osc}(\omega)$ vs. $\omega$ with Ohmic dissipation for dissipative Landau diamagnetism for different parameter values.}
\end{figure}
The quantity $S_{osc}(\omega)$ can be used as a signature for the transition from coherence to decoherence: $S_{osc}(\omega)$ has two inelastic peaks at $\omega_m=\frac{\omega_0}{2}\Big\lbrack-\kappa_2\pm\sqrt{4-\kappa_1^2+\kappa_2^2}\Big\rbrack$ for weak damping, where $\kappa_1$ and $\kappa_2$ are dimensionless parameters defined by $\kappa_1=\frac{\gamma}{\omega_0}$ and $\kappa_2=\frac{\omega_c}{\omega_0}$. These two quantities are employed as the crossover parameters to quantify the coherence to decoherence transition. Defining $\bar{\kappa}^2 = \kappa^2_1 + \kappa^2_2$, we can say that below the critical coherent criterium (defined below, c.f. Eq. (37)) i.e. $\bar{\kappa}^2<\bar{\kappa}^2_c$, the function $S_{osc}(\omega)$ exhibits two inelastic peaks which are evident from Fig. (4) in which we plot $S_{osc}(\omega)$ vs. $\omega$ for different $\kappa_1$ and $\kappa_2$. At the critical coherent criterium (c.f. Eq. (37)) the two peaks merge into a single quasielastic peak. The latter persists for $\bar{\kappa}^2>\bar{\kappa}_c^2$. Since the quasielastic peak is centered near $\omega \simeq 0$, we can make a small-$\omega$ expansion of $S_{osc}(\omega)$: 
\begin{equation}
S_{osc}(\omega) \simeq \kappa_1\chi_0^2\Big\lbrack 1 -\kappa_2\chi_0\omega+(2-\kappa_1^2-\kappa_2^2)\chi_0^2\omega^2+O(\omega^3)\Big\rbrack,
\end{equation}
where $\chi_0 = \frac{1}{\omega_0}$. The critical line is determined by inspecting the sign of the curvature of $S_{osc}(\omega)$. The latter is positive (implying coherence) if $\frac{d^2S_{osc}(\omega)}{d\omega^2}>0$, or 

\begin{equation}
\bar{\kappa}^2 = \kappa_1^2+\kappa_2^2 < 2.
\end{equation}
But the curvature changes sign at the critical line:
\begin{equation}
\bar{\kappa}_c^2 = \kappa_1^2+\kappa_2^2 = 2,
\end{equation}
and hence the system goes to the decoherent region when 
\begin{equation}
\bar{\kappa}^2 = \kappa_1^2+\kappa_2^2>2.
\end{equation}
\begin{figure}[t!]
{\rotatebox{0}{\resizebox{8cm}{8cm}{\includegraphics{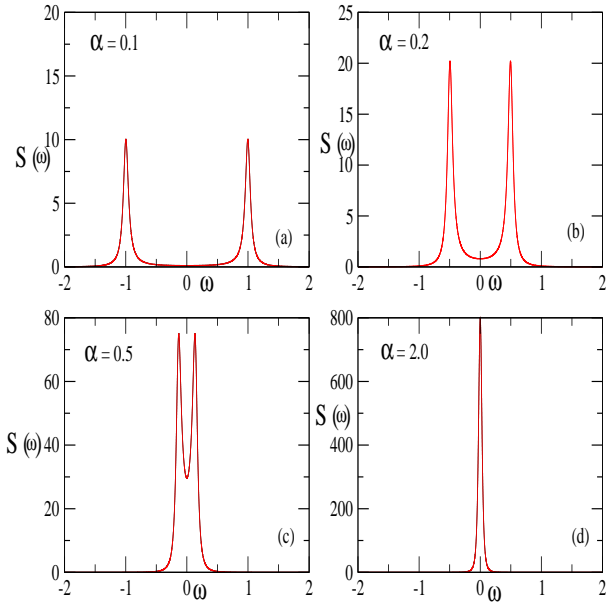}}}}
\caption{(color online) Spectral function $S_{osc}(\omega)$ vs. $\omega$ with Ohmic dissipation for a damped harmonic oscilltor for different parameter values.}\end{figure}

It is illustrative to compare this behavior with that of the damped harmonic oscillator which was discussed by Egger et al \cite{egger}. 

Comparing Figs. (4) and (5) one notes that for the damped oscillator case $S_{osc}(\omega)$ has two inelastic peaks of equal height for weak damping. As the one-parameter damping strength increases these two peaks approach each other and at the critical damping strength ($\alpha_c$) the two peaks merge into a single quasielastic peak at $\omega=0$ which persists for $\alpha>\alpha_c$. On the other hand, for  dissipative diamagnetism, the coherent-decoherent transition is to be examined in a two-parameter plane, defined by $\kappa_1$ and $\kappa_2$. One obtains two inelastic peaks which are not of equal height for low values of $\kappa_1$ and $\kappa_2$ because the peaks are not symmetric in either side of $\omega = 0$. As one increases $\kappa_1$ and $\kappa_2$ the peak height of the small peak decreases and at the critical line the small peak vanishes and we obtain a single peak which is not at $\omega = 0$, but near $\omega = 0$.  Above the critical line the single quasielastic peak persists.\\ 
We turn next to a different criterion for quantifying the transition from coherence to decoherence, which is based on the quantity $P_{osc}(t)$, defined as follows:
\begin{equation}
P_{osc}(t) = \frac{<Z(t)>}{Z_0}.
\end{equation}
We are interested in the relaxation of the expectation value $<Z(t)>$ starting from a nonequilibrium initial state. Applying the force $F(t) = m\omega_0^2Z_0$ for $t<0$, the initial condition $<Z(0)>=Z_0$ is prepared and the corresponding dynamical quantity  $P_{osc}(t)$ is computed, after switching off the force $F(t)$, at $t=0$. Following the damped quantum harmonic oscillator case \cite{egger} we may now write
\begin{equation}
P_{osc}(t) =\Re\Big\lbrack \frac{\cos(\bar{\Omega}t-\bar{\phi})\exp(-\frac{\bar{\gamma} t}{2})}{\cos(\bar{\phi})}\Big\rbrack,
\end{equation}
where
\begin{eqnarray} 
\bar{\Omega}& = &\sqrt{\omega_0^2-\frac{\bar{\gamma}^2}{4}}=\Omega^{\prime}+i\Omega^{\prime\prime}
\nonumber \\ 
\bar{\phi}& = &\Re\lbrack\tan^{-1}(\frac{\bar{\gamma}}{2\bar{\Omega}}\rbrack.
\end{eqnarray}  
Defining $a = (\omega_0^2+\frac{\omega_c^2}{4}-\frac{\gamma^2}{4})$ and $b=\frac{\gamma\omega_c}{2}$, 
\begin{eqnarray}
\Omega^{\prime}& = &\frac{1}{\sqrt{2}}\sqrt{a+\sqrt{a^2+b^2}},
\nonumber \\ 
\Omega^{\prime\prime}& = &\frac{1}{\sqrt{2}}\sqrt{\sqrt{a^2+b^2}-a},
\nonumber \\ 
\bar{\phi} & = &\tan^{-1}(X),
\nonumber \\ 
 X & = &\frac{\gamma\Omega^{\prime}+\Omega^{\prime\prime}\omega_c}{2(\Omega^{\prime 2}+\Omega^{\prime\prime}\omega_c)},
\end{eqnarray} 
\begin{widetext}
\begin{equation}
P_{osc}(t) = \Big\lbrack\frac{\cos(\Omega^{\prime}t-\bar{\phi})\cos(\Omega^{\prime\prime}t)\cos(\frac{\omega_c t}{2})-\sin(\Omega^{\prime}t-\bar{\phi})\sin(\Omega^{\prime\prime}t)\sin(\frac{\omega_c t}{2})}{\cos(\bar{\phi})}\Big\rbrack\exp(-\frac{\gamma t}{2}).
\end{equation}  
\end{widetext}
The signature of coherence is now damped-oscillatory behavior if $b^2>0$ and $a^2+b^2>0$. Thus the important inequality condition for the system to be coherent is:
\begin{equation}
(1-\kappa_1^2+\kappa_2^2)^2+\frac{(\kappa_1\kappa_2)^2}{4}>0.
\end{equation}
The system crosses over to relaxational (decoherent) behavior at the critical line
\begin{equation}
(1-\frac{\kappa_1^2}{4}+\frac{\kappa_2^2}{4})^2+\frac{(\kappa_1\kappa_2)^2}{4}=0,
\end{equation}
which is clearly different from the criterium mentioned above (cf. Eq. (37)).
Thus the criterion for crossover from coherence to decoherence depends on the specific physical quantity considered. This conclusion is similar to the cases of damped quantum harmonic oscillator as well as the spin-Boson model \cite{egger}.\\
\vskip 1cm
{\section {Summary and Conclusion}}
We have analyzed here an exact treatment of the Feynmann-Vernon model of a charged Brownian particle in a magnetic field in the quantum dissipative regime. Starting from the QLE we have derived the generalised susceptibility tensor, and have discussed its real and imaginary parts for the particular case when the magnetic field $\vec{B}$ is along the z axis. Following the Gibbs ensemble approach, we have calculated the position autocorrelation function that measures the spontaneous fluctuations of the system degrees of freedom due to coupling with the environment. The latter has been shown to be related to the imaginary part of the susceptibility that measures the energy dissipation of the system due to irreversible energy transfer between the system and the environment. The aforesaid treatment then exemplifies the fluctuation-dissipation theorem in the context of dissipative diamagnetism as well as proves the equivalence of the Einstein and the Gibbs approaches to Statistical Mechanics. Environment-induced decoherence is an important issue in  mesoscopic systems and quantum information processes. We have discussed this in the context of dissipative diamagnetism and have argued that the transition from the Landau to the Bohr-Van Leeuwen regime can indeed be viewed as a coherence to decoherence transition. Further it has been demonstrated that the initial preparation of a dissipative quantum system leads to abrupt changes regarding the criterion for coherent to decoherent transition. As in glassy systems characterized by hysteretic behavior, quantum systems also exhibit memory of their initial state of preparation. In conclusion, we have presented a unified treatment of threefold response, i.e. fluctuation, dissipation and decoherence of a system due to its coupling with environment in the context of the contemporarily important topic of dissipative diamagnetism.\\ 
\section*{Acknowledgments}
\vskip -0.55cm
 M.B. acknowledges financial support from the Council of Scientific and Industrial Research (CSIR), Government of India.

\end{document}